\documentclass{article}
\usepackage{spconf,amsmath,epsfig}

\let\OLDthebibliography\thebibliography
\renewcommand\thebibliography[1]{
  \OLDthebibliography{#1}
  \setlength{\parskip}{0pt}
  \setlength{\itemsep}{0pt plus 0.3ex}
}

\usepackage{epsfig}
\usepackage{graphicx}
\usepackage{amsmath}
\usepackage{amssymb}
\usepackage{booktabs}
\usepackage{graphicx}
\usepackage{subfigure}
\usepackage{multirow}
\usepackage{bm, amsmath}
\usepackage{makecell}
\usepackage{color}
\usepackage{rotating}
\usepackage{enumitem}
\usepackage{makecell}
\usepackage{newfloat}
\usepackage{listings}

\begin{document}\sloppy

\def\x{{\mathbf x}}
\def\L{{\cal L}}

\title{Multi-perspective Memory Enhanced Network for Identifying Key Nodes in Social Networks}
%
\name{Qiang Zhang\textsuperscript{\rm 1}, Jiawei Liu\textsuperscript{\rm 1}, Fanrui Zhang\textsuperscript{\rm 1},  Xiaoling Zhu\textsuperscript{\rm 2}, Zheng-Jun Zha\textsuperscript{\rm 1}}
\address{\textsuperscript{\rm 1} University of Science and Technology of China, China\\
                    \textsuperscript{\rm 2} CETC Academy of Electronics and Information Technology Group Co, China\\
                    \{zq\_126, zfr888\}@mail.ustc.edu.cn\\
                    zhuxiaoling1@cetc.com.cn\\
                    \{jwliu6, zhazj\}@ustc.edu.cn
                    }

\maketitle

\begin{abstract}
Identifying key nodes in social networks plays a crucial role in timely blocking false information. Existing key node identification methods usually consider node influence only from the propagation structure perspective and have insufficient generalization ability to unknown scenarios. In this paper, we propose a novel Multi-perspective Memory Enhanced Network (MMEN) for identifying key nodes in social networks, which mines key nodes from multiple perspectives and utilizes memory networks to store historical information. Specifically, MMEN first constructs two propagation networks from the perspectives of user attributes and propagation structure and updates node feature representations using graph attention networks. Meanwhile, the memory network is employed to store information of similar subgraphs, enhancing the model's generalization performance in unknown scenarios. Finally, MMEN applies adaptive weights to combine the node influence of the two propagation networks to select the ultimate key nodes. Extensive experiments demonstrate that our method significantly outperforms previous methods.
\end{abstract}
\section{Introduction}
\label{sec:intro}

As social media platforms increasingly integrate into people's lives, they have become the primary source of information for the public. On social platforms, users can freely share and disseminate content, but their high efficiency and fast-paced propagation capabilities also provide conditions for the generation and spread of false information \cite{mishra2020fake}. Due to the chaotic nature of false information, people are often misled, affecting their judgment and decision-making. The false information can also be used to distort and fabricate facts to manipulate public opinion, adversely impacting social trust and stability \cite{li2023edge,narayan2022desi}. Therefore, identifying key nodes in news propagation networks to timely block false information has attracted significant attention from the research community \cite{zhang2019discount,zhou2019fast}.

Modeling node influence in propagation networks and identifying key nodes can indirectly grasp the mechanism of content dissemination, thereby assisting in formulating strategies to block false media content. So far, many methods have been proposed to measure the importance of nodes in propagation networks. Among them, degree centrality \cite{bonacich1972factoring}, K-shell \cite{kitsak2010identification}, H-index \cite{lu2016h,pastor2017topological}, and semi-local centrality \cite{chen2012identifying} are based on subgraphs constituted by the node itself and its neighboring nodes to select influential nodes. HIT \cite{kleinberg1999authoritative}, LeaderRank \cite{lu2011leaders}, and PageRank \cite{horowitz2010anatomy} methods are based on node feature vectors for key node identification. With the development of machine learning, some researchers have proposed the concept of Graph Convolutional Network (GCN) \cite{hamilton2017representation}, aiming to learn the low-dimensional latent representation of graphs and perform various tasks in complex networks. Therefore, some methods apply GCN to transform the key node identification problem in complex networks into the regression problem, but these methods only model node influence based on the network structure perspective \cite{zhao2020infgcn,yu2020identifying,fan2020finding}, lacking the analysis of node influence from the user attributes themselves. Moreover, these methods lack the preservation of past structural information, limiting the model's generalization ability in different scenarios. Due to the fact that different nodes and their adjacent subgraphs may have similar structures, and consequently analogous influence patterns, employing memory networks to store information of similar subgraphs and retrieve information from them can effectively explore and utilize the similarities between different subgraphs, thereby improving the network's generalization performance.

In this work, we propose a novel Multi-perspective Memory Enhanced Network (MMEN) for identifying key nodes in social networks, which mines key nodes from multiple perspectives and utilizes memory networks to store historical information. Specifically, the proposed MMEN is divided into three modules: social graph construction module, graph memory enhancement module, and multi-perspective fusion module. First, for the social graph construction module, we model the propagation network from both user attributes and propagation structure perspectives into two isomorphic propagation networks. The user attribute network embedding encodes numerical vectors based on multiple user attribute information, while the propagation structure network uses the random walk strategy \cite{nikolentzos2020random} to sample a fixed number of neighboring nodes for each node, thus obtaining the initial node feature representation.

For the graph memory enhancement module, we apply the graph attention network \cite{velivckovic2017graph} to update the node's feature representation, and to leverage the feature similarity between different subgraphs, we store each subgraph's information in the memory module. When predicting, the nodes queries similar subgraph features from the memory module and connects the memory features with the initial features of the node to be predicted. In the multi-perspective fusion module, we utilize adaptive weights to aggregate the influence prediction scores of the nodes in the two isomorphic propagation networks and ultimately select key nodes with a greater impact on information dissemination. Extensive experiments on two datasets demonstrate that our method outperforms others key node identification approaches.

In summary, the main contributions of this paper are described below:
\begin{itemize}
    \item We propose a novel multi-perspective memory enhanced network for identifying key nodes in social networks, which mines key nodes from multiple perspectives and utilizes memory networks to store historical subgraph information.
    \item We utilize memory networks to store and retrieve information of similar subgraphs, exploring the use of similarity between different subgraphs, which can effectively enhance the network's generalization performance in unknown scenarios..
    \item We mine key nodes from both user attributes and propagation structure perspectives, and apply adaptive weights to aggregate the influence prediction scores of the nodes in the two isomorphic propagation graphs, ultimately selecting the key nodes.
\end{itemize}

\begin{figure*}[htbp]
	\centering
    \includegraphics[width=1\textwidth]{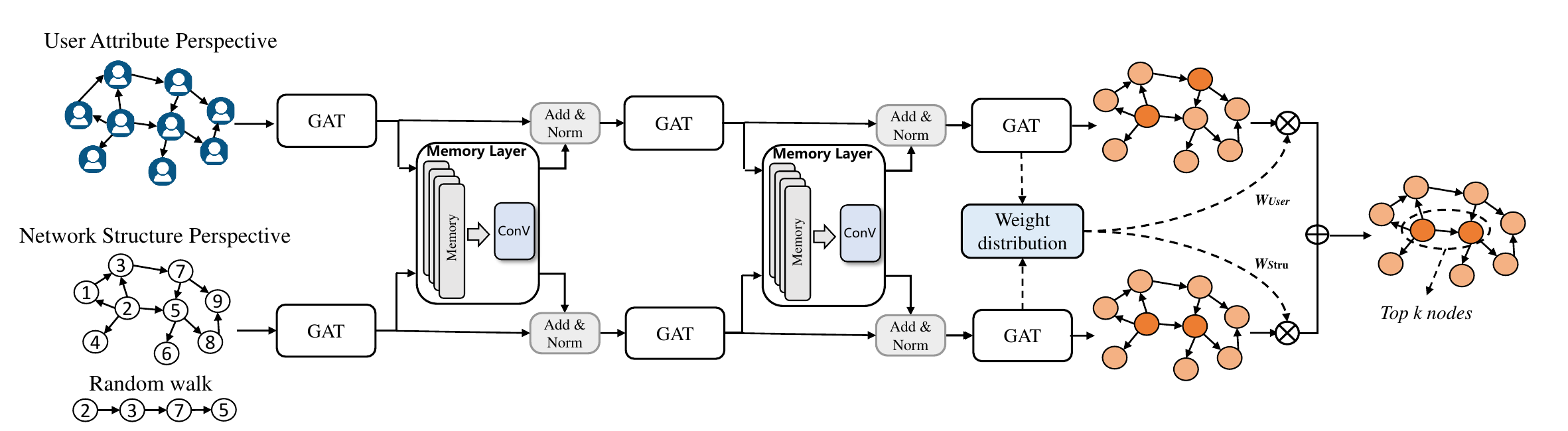}
	\caption{The overall architecture of the proposed MMEN. It consists of three modules: social graph construction module, graph memory enhancement module and multi-perspective fusion module.}
	\label{fig:model}
\end{figure*}

\section{Method}
\subsection{Model Overview}
In this section, we will elaborate on our method MMEN. The MMEN's architecture is showed in Figure \ref{fig:model}. Specifically, MMEN consists of three modules for extracting node features and performing key node identification: (1) Social graph construction module, which models the propagation network from both user attributes and propagation structure perspectives into two isomorphic networks. (2) Graph memory enhancement module, which utilizes the graph attention network to update the node's feature representation and perform feature enhancement, then stores each subgraph's information in the memory module to improve the model's generalization performance. (3) Multi-perspective fusion module, in which we apply adaptive weights to aggregate the influence prediction scores of the nodes in the two isomorphic networks and ultimately select key nodes. In the following, we will provide the detailed introduction to these modules.

\subsection{Problem Definition}
The propagation network can be defined as $G=\left( V,E \right) $, where $V=\left\{ v_1,v_2,...,v_n \right\} $ represents the set of users in the social network, and $E\subseteq V\times V$ represents the retweeting relationships between users in the social network. If there is a retweeting relationship from node $v_i$ to node $v_j$, then $E_{i,j}=1$; otherwise, $E_{i,j}=0$. $d\left( v_i,v_j \right) $ represents the distance between node $v_i$ and node $v_j$, that is, the shortest path length between them. The task objective of key node identification is to find the top 5\% most influential nodes in the entire propagation network.

\subsection{Social Graph Construction}
For a propagation event graph $G$, based on its source post and retweeting, we construct two isomorphic propagation graphs: one is the user attribute graph $G_u$, with node embeddings based on user attribute vectors, and the other is the propagation structure graph $G_s$, with node embeddings based on the random walk algorithm to extract structural information around the nodes. The construction processes are described separately as follows:

\textbf{User attribute graph.} To describe how users participate in social networks, we apply their metadata and profiles to define the feature vector $x_i$ for each user $u_i$, which serves as the embedding for the user attribute graph. The extracted features include: (1) the number of characters in $u_i$'s name, (2) the number of characters in $u_i$'s self-description, (3) the number of other users following $u_i$, (4) the number of users $u_i$ follows, (5) the number of tweets created by $u_i$, (6) whether $u_i$'s account is verified, (7) whether $u_i$ allows geospatial location, (8) the time difference between the source tweet's release time and $u_i$'s retweeting time, and (9) the retweeting path length between $u_i$ and the source tweet. We map all user attribute information to numerical vectors and combine these nine features in sequence to obtain the feature vector $x_i$ for each user.

\textbf{Propagation structure graph.} For the embeddings in the propagation structure graph, we utilize the random walk \cite{nikolentzos2020random} strategy to sample a fixed number of neighboring nodes for each node, obtaining the corresponding subgraph and initial feature representation.

\subsection{Graph Memory Enhancement}
\textbf{GAT network.} Based on the two types of propagation networks described above, we apply the GAT network \cite{velivckovic2017graph} to adaptively calculate the attention weights between nodes to capture information in the graph structure. Based on this module, we first calculate the attention coefficients for corresponding nodes to their neighboring nodes, and use the \textit{softmax} function to normalize them, obtaining the final attention scores $a_{i}$. The calculation formula is described as follows:
\begin{equation}
    a_{i}=soft\max \left( \sigma \left( \overrightarrow{a}^T\left[ W\vec{h}_i||W\vec{h}_j \right] \right) \right), j \in \mathcal{N}_{i}
\end{equation}

where $h_i$ represents the central node of the computation, $h_j$ represents the neighbouring nodes and $W$ represents a linear layer with shared parameters

Based on the obtained attention scores $a_{i}$, we perform a weighted summation of the neighboring node features to obtain the enhanced center node features.
\begin{equation}
    h_{i}^{\prime}\left( K \right) =\underset{k=1}{\overset{K}{||}}\sigma \left( \sum{_{j\in N_i}a_{i}^{k}W^kh_j} \right) 
\end{equation}

Where $K$ represents the number of attention heads of the GAT network.

\textbf{Memory network module.} Nodes and their adjacent subgraphs in different networks are likely to have similar subgraph structures and exhibit comparable influence patterns. To enhance the model's generalization ability in unknown scenarios, we employ the memory network module to model this structural similarity. By storing information of similar subgraphs in the memory network and retrieving it, we can effectively exploit the similarities between different subgraphs.

Specifically, we designed multiple groups of memory layers \cite{sukhbaatar2015end} $M=\left\{ m_1,m_2,...,m_n \right\} $ to store network structure information, where $n$ represents the number of memory layers, and the dimension size of each group of storage networks is $m_i\in \mathbb{R} ^{b\times L}$, $b$ represents the number of memory storage variables. For each data sample, the node features enhanced by the GAT network will find the closest similar structure information from each memory layer and extract the corresponding feature representation.

The computation steps are as follows: Given the network node features $h$, we compute the similarity probabilities between the node features and each memory information by applying the $softmax$ function. Then, based on these similarity probabilities, we perform a weighted sum of each memory information to obtain the latent structural features $F_{mi}$ of the network. The formula is as follows:

\begin{equation}
F_{mi}=soft\max \left( h^Tm_i \right)m_i 
\end{equation}

After obtaining the potential structural features of the multi-group memory layers, we apply a 1D convolutional network to aggregate the information of different memory layers to obtain the final memory information $F_m$.

Finally, we connect the resulting features $F_m$ with the original node features to obtain the memory-enhanced network structure information $F_a$.

\begin{equation}
F_a=\mathrm{Re}lu\left( layernorm\left( h+F_m \right) \right) 
\end{equation}

The above is the calculation process of a single-layer graph memory enhancement module. After calculating the two propagation graphs with the two-layer graph memory enhancement module, we employ a fully connected layer to map the features of each node to the range of [0, 1], which represents the node influence score $S_i(v),i\in (1,2)\ and\ v\in V$ obtained from corresponding propagation graph. The computation formula is shown below:

\begin{equation}
S_i=soft\max \left( W_s\left( R \right) +b_s \right) 
\end{equation}

\subsection{Multi-Perspective Fusion}

In order to aggregate the node influence scores $S_1$ from the perspective of user attributes and $S_2$ from the perspective of propagation structure, we design an adaptive weight assignment module to aggregate them and get the final node influence scores. The specific process is described as follows:

First, based on the graph embedding features $h_u$ extracted from the user attribute graph and the features $h_s$ extracted from the propagation structure graph, we apply average pooling operations to compress them into two one-dimensional vectors. Then, we concatenation these two one-dimensional vectors and feed them into a fully connected layer, normalizing them to the range of [0-1] using the \textit{Softmax} function, thus obtaining the final weight distribution $w_{user}$ and $w_{stru}$. The calculation formula is shown below:

\begin{equation}
w=soft\max \left( W_m\left( mean\left( h_u \right) \oplus mean\left( h_s \right) \right) +b_m \right) 
\end{equation}

Finally, we multiply the node influence scores of the two propagation graphs by $w_{user}$ and $w_{stru}$, respectively, and perform a summation process to obtain the final aggregated influence score $S$.

\begin{equation}
S=m_{user}\times S_1+m_{stru}\times S_2
\end{equation}

\subsection{Model Optimization}
Inspired by the paper \cite{ni2021fastcover}, we choose to interpret the scores in a probabilistic manner, thus optimizing the entire network using unsupervised algorithms. We want the scores $s_v,\ v\in V$ to represent the probability of each node $v$ being selected as part of the seed set. The loss function of the scores p of the nodes in graph $G$ is defined as follows:

\begin{equation}
\mathcal{L}(\mathbf{s}, G) \triangleq \mathbb{E}[\mid \text{uncovered vertices}\mid]+\lambda \mathbb{E}[\mid\text{seed set}\mid]
\end{equation}

where $\lambda > 0$, and the expectation is calculated over the randomness of selecting the seed set by integrating $s_v$. The first term in penalizes the expected number of unaffected vertices. Minimizing this term is equivalent to maximizing the number of vertices covered. The second term regularizes the expected size of the seed set so that vertices with lower importance are less likely to be selected. Without the latter term, the optimal solution for  would obviously be $s_v = 1$ for all $v\in V$. $\lambda$ balances the influence of these two terms.

Thus, the loss function can be explicitly written as:

\begin{equation}
\mathcal{L}(\mathbf{s}, G)=\sum_{v \in V} \prod_{v: u \in \mathcal{N}_{d}^{+}(v)}\left(1-s_{v}\right)+\lambda \sum_{v \in V} s_{v}
\end{equation}

\section{EXPERIMENTS}
We evaluate the performance of MMEN through the SIR propagation model and the network robustness index R, and compare it with well-known existing indicators in two real social media propagation networks.

\subsection{Experimental Settings}
\subsubsection{Datasets} 
We utilize two famous datasets Twitter15 and Twitter16 \cite{ma2017detect} to validate our model performance. Each dataset contains a collection of source tweets and the corresponding retweeting user sequences. Since the original data does not contain user profiles, we crawl user information using the Twitter API with user IDs.

\subsubsection{Evaluation Metrics}
\textbf{Node infection rate $\bm{S_t}$.}
In the SIR model, each node is in one of three states: susceptible (S), infected (I), and recovered (R). Susceptible nodes can be infected at each time step. Infected nodes have been infected and attempt to infect susceptible nodes among their neighbors at each time step with probability $\mu$. Recovered nodes have recovered from the infected state and will not be infected by infected nodes again. At each time step, infected nodes will recover with probability $\beta$ (for simplicity, $\beta = 1$ in this paper). The process terminates if there are no infected nodes in the network. We can set a set of initial nodes as infected nodes and other nodes as susceptible nodes to estimate the influence of that set of nodes nodes in the network. In this paper, $S_t$ is defined as the ratio of infected nodes when the propagation process reaches a steady state.

\textbf{Network robustness index $\bm{R}$.} To investigate the importance of the selected node group for network connectivity, we remove this group of nodes to calculate the network robustness index $R$. The approach is to calculate the proportion of the remaining main network nodes to the original network after removing the selected node group.
\begin{equation}
R=Z/N
\end{equation}
where $Z$ represents the number of nodes in the maximum connected subset of the network after removing a portion of nodes, $N$ represents the total number of nodes in the network.

\subsubsection{Implementation Details} 
While crawling the user information in the dataset, if some attribute information of the user is not available, its value will be replaced with 0. In graph memory enhancement module, each memory network has 4 layers. For parameter settings, we set \textit{Batch size = 2}, \textit{Epoch = 50}, \textit{learning rate = 5e-4}, the optimizer is Adam, and we utilize early stopping strategy.

\begin{table*}[!t]
\caption{ Performance comparison to the other methods on Twitter15 and Twitter16 datasets.}
  \centering
  \footnotesize
  \resizebox{1.5\columnwidth}{!}{
  \renewcommand{\arraystretch}{1.0}
  \begin{tabular}{c|cc|cc|cc}
    \hline
    \multirow{2}{*}{\textbf{Method}} & \multicolumn{2}{c|}{\textbf{Perspective}} & \multicolumn{2}{c|}{\textbf{Twitter15}} & \multicolumn{2}{c}{\textbf{Twitter16}} \\
    \cline{2-7}
    &Structure&User&Infection rate $S_t$& $R$ &Infection rate $S_t$& $R$\\
    \hline
    K-Shell \cite{kitsak2010identification} & \checkmark & & 0.534 & 0.487  & 0.428 & 0.563\\
    LeaderRank \cite{lu2011leaders} & \checkmark & & 0.649 & 0.384 & 0.587 & 0.495 \\
    Greedy \cite{basuchowdhuri2014finding} & \checkmark & & 0.683 & 0.364 & 0.572 & 0.513\\
    HEU \cite{nguyen2020solving} & \checkmark & & 0.786 & 0.189 & 0.732 & 0.281 \\
    S2V-DQN \cite{khalil2017learning} & \checkmark & & 0.792 & 0.192 & 0.721 & 0.289\\
    Fastcover \cite{ni2021fastcover} & \checkmark & & 0.834 & 0.174 & 0.761 & 0.261\\
    \textbf{MMEN} & \checkmark & \checkmark & \textbf{0.876} & \textbf{0.128} & \textbf{0.809} & \textbf{0.234}\\
    \hline
  \end{tabular}
  }
\label{tab:comparation}
\end{table*}

\subsection{Baselines}
The comparison methods are described as follows: 
\begin{itemize}
    \item K-Shell \cite{kitsak2010identification}: By layering the nodes in the network according to their connectivity, forming a series of shells to understand the network structure and identify important nodes in the network.
    \item LeaderRank \cite{lu2011leaders}: determines the influence of network nodes by adding virtual nodes to the network and thus considering the in-degree and out-degree of the nodes.
    \item Greedy \cite{basuchowdhuri2014finding}: In each iteration, Greedy adds the node with the maximum effective d coverage to the seed set and uses breadth-first search (BFS) to update the effective coverage.
    \item HEU \cite{nguyen2020solving}: HEU is a lightweight three-phase algorithm that integrates various heuristic algorithms based on features observed in real graphs.
    \item S2V-DQN \cite{khalil2017learning}: S2V-DQN is a framework for solving graph combination problems through reinforcement learning. It uses multilayer GNN for graph and node embedding, and learns a greedy strategy for iterative node selection based on DQN.
    \item Fastcover \cite{ni2021fastcover}: Discovering influential nodes efficiently with unsupervised learning through attention-based graph neural networks.
\end{itemize}

\subsection{Results and Discussion}
The experimental results are shown in Table \ref{tab:comparation}. Under both evaluation metrics, MMEN achieves the optimal performance, with the infection rate $S_t$ of 87.6\% and $R$ index of 12.8\% in the Twitter15 dataset, and the infection rate $S_t$ of 80.9\% and $R$ index of 23.4\% in the Twitter16 dataset, reflecting the excellent performance of MMEN.

Among them, methods such as FastCover and S2V-DQN utilize GNN for node embedding and update node features, their performance is better than traditional methods, indicating that networks like GCN can effectively extract network structure features and benefit the key node identification task. Compared with other methods, our MMEN model outperforms other models, which can attribute the advantages of the MMEN model to the following factors: (1) Mining key nodes from both user attributes and propagation structure perspectives, and using adaptive weights to integrate information from various viewpoints. (2) Using memory networks to store and retrieve information on similar subgraphs, thereby enhancing the network's generalization performance.

\begin{table}[!t]
	\caption{Ablation study of different components of MMEN on the Twitter15 dataset.}
	\begin{center}
		\begin{tabular}{c|cc}
			\hline
			\multirow{2}{*}{\textbf{Method}} & \multicolumn{2}{c}{\textbf{Twitter15}} \\
            \cline{2-3}
             &Infection rate $S_t$&$R$\\
			\hline
             w/o User attribute & 0.847 & 0.183 \\
             w/o Memory layer & 0.861 & 0.142 \\
             w/o Adaptive Fusion & 0.869 & 0.134 \\
             \textbf{MMEN} & \textbf{0.876} & \textbf{0.128}  \\
			\hline		
		\end{tabular}
		\label{tab: Ablation Study}
	\end{center}
\end{table}

\subsection{Ablation Studies}

The ablation experiments in Table \ref{tab: Ablation Study} investigate the impact of each module component of MMEN on the key node identification performance. Specifically, the MMEN variants are described as follows: \textit{w/o User attribute} refers to not using the user attribute graph, \textit{i.e.}, only applying the propagation structure graph to identify key nodes. \textit{w/o Memory layer} refers to not using the memory network to store information on similar subgraphs. \textit{w/o Adaptive Fusion} refers to not using the adaptive weight allocation module to aggregate the node influence of the two subgraphs, and their influences will be directly summed up to select the key nodes. The performance of these MMEN variants is significantly worse than the original MMEN, with \textit{w/o User attribute} having the worst performance among these variants. This indicates that: (1) The user attribute graph can effectively supplement user-level information and significantly improve model performance. (2) Using the memory network module can enhance the network's generalization performance when facing unknown subgraphs. (3) The multi-perspective fusion module can combine node influence from both user and propagation structure perspectives, thereby identifying key nodes from multiple angles.

\section{CONCLUSION}
In this paper, we propose a novel multi-perspective memory enhanced network for identifying key nodes in social networks, which mines key nodes from multiple perspectives and utilizes memory networks to store historical information. The MMEN model adaptively integrates both perspectives of user attributes and propagation structure to identify key nodes, and enhance the generalisation performance of the model in unknown scenarios through the memory network. Extensive experiments demonstrate that our proposed model MMEN significantly outperforms previous approaches and accurately identifies influential nodes.

\bibliographystyle{IEEEbib}
\bibliography{icme2023template}

\begin{thebibliography}{10}

\bibitem{Authors12}
Authors,
\newblock ``The frobnicatable foo filter,'' ACM MM 2013 submission ID 324. Supplied as additional material {\tt acmmm13.pdf}.

\bibitem{Authors12b}
Authors,
\newblock ``Frobnication tutorial,'' 2012,
\newblock Supplied as additional material {\tt tr.pdf}.

\bibitem{cooley65}
J.~W. Cooley and J.~W. Tukey,
\newblock ``An algorithm for the machine computation of complex {F}ourier series,''
\newblock {\em Math. Comp.}, vol. 19, pp. 297--301, Apr. 1965.

\bibitem{haykin02}
S.~Haykin,
\newblock ``Adaptive filter theory,''
\newblock Information and System Science. Prentice Hall, 4th edition, 2002.

\bibitem{Morgan2005}
Dennis~R. Morgan,
\newblock ``Dos and don'ts of technical writing,''
\newblock {\em IEEE Potentials}, vol. 24, no. 3, pp. 22--25, Aug. 2005.

\bibitem{huh2018fighting}
Minyoung Huh, Andrew Liu, Andrew Owens, and Alexei~A Efros,
\newblock ``Fighting fake news: Image splice detection via learned self-consistency,''
\newblock in {\em Proceedings of the European conference on computer vision (ECCV)}, 2018, pp. 101--117.

\bibitem{mishra2020fake}
Rahul Mishra,
\newblock ``Fake news detection using higher-order user to user mutual-attention progression in propagation paths,''
\newblock in {\em Proceedings of the IEEE/CVF conference on computer vision and pattern recognition workshops}, 2020, pp. 652--653.

\bibitem{li2023edge}
Dong Li, Jiaying Zhu, Menglu Wang, Jiawei Liu, Xueyang Fu, and Zheng-Jun Zha,
\newblock ``Edge-aware regional message passing controller for image forgery localization,''
\newblock in {\em Proceedings of the IEEE/CVF Conference on Computer Vision and Pattern Recognition}, 2023, pp. 8222--8232.

\bibitem{narayan2022desi}
Kartik Narayan, Harsh Agarwal, Surbhi Mittal, Kartik Thakral, Suman Kundu, Mayank Vatsa, and Richa Singh,
\newblock ``Desi: Deepfake source identifier for social media,''
\newblock in {\em Proceedings of the IEEE/CVF Conference on Computer Vision and Pattern Recognition}, 2022, pp. 2858--2867.

\bibitem{zhang2019discount}
Tianrui Zhang, Pengdeng Li, Lu-Xing Yang, Xiaofan Yang, Yuan~Yan Tang, and Yingbo Wu,
\newblock ``A discount strategy in word-of-mouth marketing,''
\newblock {\em Communications in Nonlinear Science and Numerical Simulation}, vol. 74, pp. 167--179, 2019.

\bibitem{zhou2019fast}
Fang Zhou, Linyuan L{\"u}, and Manuel~Sebastian Mariani,
\newblock ``Fast influencers in complex networks,''
\newblock {\em Communications in Nonlinear Science and Numerical Simulation}, vol. 74, pp. 69--83, 2019.

\bibitem{bonacich1972factoring}
Phillip Bonacich,
\newblock ``Factoring and weighting approaches to status scores and clique identification,''
\newblock {\em Journal of mathematical sociology}, vol. 2, no. 1, pp. 113--120, 1972.

\bibitem{kitsak2010identification}
Maksim Kitsak, Lazaros~K Gallos, Shlomo Havlin, Fredrik Liljeros, Lev Muchnik, H~Eugene Stanley, and Hern{\'a}n~A Makse,
\newblock ``Identification of influential spreaders in complex networks,''
\newblock {\em Nature physics}, vol. 6, no. 11, pp. 888--893, 2010.

\bibitem{lu2016h}
Linyuan L{\"u}, Tao Zhou, Qian-Ming Zhang, and H~Eugene Stanley,
\newblock ``The h-index of a network node and its relation to degree and coreness,''
\newblock {\em Nature communications}, vol. 7, no. 1, pp. 10168, 2016.

\bibitem{pastor2017topological}
Romualdo Pastor-Satorras and Claudio Castellano,
\newblock ``Topological structure and the h index in complex networks,''
\newblock {\em Physical Review E}, vol. 95, no. 2, pp. 022301, 2017.

\bibitem{kleinberg1999authoritative}
Jon~M Kleinberg,
\newblock ``Authoritative sources in a hyperlinked environment,''
\newblock {\em Journal of the ACM (JACM)}, vol. 46, no. 5, pp. 604--632, 1999.

\bibitem{lu2011leaders}
Linyuan L{\"u}, Yi-Cheng Zhang, Chi~Ho Yeung, and Tao Zhou,
\newblock ``Leaders in social networks, the delicious case,''
\newblock {\em PloS one}, vol. 6, no. 6, pp. e21202, 2011.

\bibitem{horowitz2010anatomy}
Damon Horowitz and Sepandar~D Kamvar,
\newblock ``The anatomy of a large-scale social search engine,''
\newblock in {\em Proceedings of the 19th international conference on World wide web}, 2010, pp. 431--440.

\bibitem{hamilton2017representation}
William~L Hamilton, Rex Ying, and Jure Leskovec,
\newblock ``Representation learning on graphs: Methods and applications,''
\newblock {\em arXiv preprint arXiv:1709.05584}, 2017.

\bibitem{zhao2020infgcn}
Gouheng Zhao, Peng Jia, Anmin Zhou, and Bing Zhang,
\newblock ``Infgcn: Identifying influential nodes in complex networks with graph convolutional networks,''
\newblock {\em Neurocomputing}, vol. 414, pp. 18--26, 2020.

\bibitem{yu2020identifying}
En-Yu Yu, Yue-Ping Wang, Yan Fu, Duan-Bing Chen, and Mei Xie,
\newblock ``Identifying critical nodes in complex networks via graph convolutional networks,''
\newblock {\em Knowledge-Based Systems}, vol. 198, pp. 105893, 2020.

\bibitem{nikolentzos2020random}
Giannis Nikolentzos and Michalis Vazirgiannis,
\newblock ``Random walk graph neural networks,''
\newblock {\em Advances in Neural Information Processing Systems}, vol. 33, pp. 16211--16222, 2020.

\bibitem{velivckovic2017graph}
Petar Veli{\v{c}}kovi{\'c}, Guillem Cucurull, Arantxa Casanova, Adriana Romero, Pietro Lio, and Yoshua Bengio,
\newblock ``Graph attention networks,''
\newblock {\em arXiv preprint arXiv:1710.10903}, 2017.

\bibitem{sukhbaatar2015end}
Sainbayar Sukhbaatar, Jason Weston, Rob Fergus, et~al.,
\newblock ``End-to-end memory networks,''
\newblock {\em Advances in neural information processing systems}, vol. 28, 2015.

\bibitem{hu2018squeeze}
Jie Hu, Li~Shen, and Gang Sun,
\newblock ``Squeeze-and-excitation networks,''
\newblock in {\em Proceedings of the IEEE conference on computer vision and pattern recognition}, 2018, pp. 7132--7141.

\bibitem{ni2021fastcover}
Runbo Ni, Xueyan Li, Fangqi Li, Xiaofeng Gao, and Guihai Chen,
\newblock ``Fastcover: An unsupervised learning framework for multi-hop influence maximization in social networks,''
\newblock {\em arXiv preprint arXiv:2111.00463}, 2021.

\bibitem{basuchowdhuri2014finding}
Partha Basuchowdhuri and Subhashis Majumder,
\newblock ``Finding influential nodes in social networks using minimum k-hop dominating set,''
\newblock in {\em Applied Algorithms: First International Conference, ICAA 2014, Kolkata, India, January 13-15, 2014. Proceedings 1}. Springer, 2014, pp. 137--151.

\bibitem{nguyen2020solving}
Minh~Hai Nguyen, Minh~Ho{\`a}ng H{\`a}, Diep~N Nguyen, and The~Trung Tran,
\newblock ``Solving the k-dominating set problem on very large-scale networks,''
\newblock {\em Computational Social Networks}, vol. 7, pp. 1--15, 2020.

\bibitem{khalil2017learning}
Elias Khalil, Hanjun Dai, Yuyu Zhang, Bistra Dilkina, and Le~Song,
\newblock ``Learning combinatorial optimization algorithms over graphs,''
\newblock {\em Advances in neural information processing systems}, vol. 30, 2017.

\bibitem{chen2012identifying}
Duanbing Chen, Linyuan L{\"u}, Ming-Sheng Shang, Yi-Cheng Zhang, and Tao Zhou,
\newblock ``Identifying influential nodes in complex networks,''
\newblock {\em Physica a: Statistical mechanics and its applications}, vol. 391, no. 4, pp. 1777--1787, 2012.

\bibitem{ma2017detect}
Jing Ma, Wei Gao, and Kam-Fai Wong,
\newblock ``Detect rumors in microblog posts using propagation structure via kernel learning,''
\newblock Association for Computational Linguistics, 2017.

\bibitem{fan2020finding}
Changjun Fan, Li~Zeng, Yizhou Sun, and Yang-Yu Liu,
\newblock ``Finding key players in complex networks through deep reinforcement learning,''
\newblock {\em Nature machine intelligence}, vol. 2, no. 6, pp. 317--324, 2020.

\end{thebibliography}

\end{document}